\documentclass[twocolumn]{jpsj2} 
\title{
Basal-Plane Magnetic Anisotropies of High-$\kappa$ $d$-Wave Superconductors\\ 
in a Mixed State: A Quasiclassical Approach}

\author{Hiroto \textsc{ADACHI}\thanks{E-mail address: adachi@mp.okayama-u.ac.jp}, 
Predrag \textsc{Miranovi\'{c}}$^{1}$, 
Masanori \textsc{ICHIOKA} 
and Kazushige \textsc{MACHIDA} }

\inst{
Department of Physics, Okayama University, Okayama 700-8530 \\
$^{1}$
Department of Physics, University of Montenegro, Podgorica 81000, 
Serbia and Montenegro
}

\abst{
We study the basal-plane anisotropies of reversible magnetization and torque 
in a mixed state of layered $d$-wave superconductors 
based on the quasiclassical version of the BCS-Gor'kov theory. 
Both the longitudinal magnetization ($M_L$) and torque ($\tau$) 
show fourfold oscillations as a function of the field angle $\chi$. 
The relationship between the node position and 
the oscillatory patterns shown by $M_L$ 
and $\tau$ is clarified. 
It is also shown that the sign of the $\tau (\chi)$-oscillation does not change
between $H_{c1}$ and $H_{c2}$, 
while the sign of the $M_L (\chi)$-oscillation changes. 
The newly obtained result for $\tau$ indicates that the torque experiment can 
allow us to detect 
the in-plane anisotropies of $H_{c2}$ even in a material with strong fluctuations 
such as cuprate or organic superconductors, where the $H_{c2}$ itself cannot be 
determined experimentally. 
}

\kword{$d$-wave superconductor, mixed state, magnetization, torque, in-plane anisotropy} 

\begin{document}
\maketitle

\section{Introduction}
It has been widely known that the effects of nonlocal electrodynamics 
in a clean type-II superconductor cause several macroscopic anisotropy 
phenomena.~\cite{Weber} 
  For the uniaxial anisotropy in a layered superconductor, or the basal-plane 
anisotropy in a multicomponent superconductor, 
the local London~\cite{Kogan88} 
or the local Ginzburg-Landau~\cite{Hao, Machida85} (GL) 
approach provides a simple and convenient description of the phenomenon 
in terms of the second rank mass tensor. 
However, as pointed out by Hohenberg and Werthamer,~\cite{HW} 
the anisotropy in a cubic system 
cannot be explained within the framework of these local theories, 
and in such a case, we must recourse to theories incorporating 
nonlocal effects.~\cite{Kogan96, Kosztin, Franz, Takanaka, Ichioka} 
Among these nonlocal theories, 
the quasiclassical Eilenberger formalism~\cite{Ichioka, Eilenberger, LO} 
is advantageous for a quantitative description of the phenomena. 
It can also be employed to interpolate the results obtained using 
the London theory and the GL theory, 
valid at $H \ll H_{c2}$ and $H \approx H_{c2}$, respectively, 
since both theories can be derived from the Eilenberger formalism. 

Recently, in the field of unconventional superconductivity, 
there has been renewed interest in such a mixed-state anisotropy phenomenon. 
Typical examples are the field-angle-dependent oscillations 
of the specific heat~\cite{Park, Aoki} and thermal conductivity.~\cite{Izawa} 
These phenomena can be used to clarify the node position of the gap function, 
through the fact that the Doppler-shifted quasiparticles cause the 
variation of these quantities 
depending on the angle between the applied field direction 
and the nodal direction. 
For more details, we refer the reader to the seminal theoretical work 
by Vekhter {\it et al.}~\cite{Vekhter} 
and a more quantitative calculation by Miranovi\'{c} {\it et al}.~\cite{Pedja1} 

Magnetization is another fundamental quantity in the mixed state of 
a type-II superconductor. 
Thanks to the newly developed ``shaking'' technique~\cite{Willemin} and 
the arrival of high-quality low-pinning samples, 
we can now obtain reversible magnetizations in a broad 
region of the field-temperature phase diagram. 
Indeed there exist precise reversible magnetization 
measurements~\cite{Civale1, Kogan99}  
for borocarbides, showing clear fourfold basal-plane anisotropies. 
In our previous work,~\cite{Adachi1} we studied 
such a basal-plane anisotropy of the longitudinal magnetization $M_L$. 
We demonstrated that the experimental data of $M_L$ for borocarbides 
can be explained by considering both gap and Fermi surface anisotropies. 
However, results on the transverse component $M_T$ of the reversible 
magnetization, or torque $\tau$, have not been discussed. 

Experimentally, the basal-plane magnetic torque 
in a mixed state 
was measured by Ishida {\it et al.}~\cite{Ishida} for an 
untwinned single crystal, YBa$_2$Cu$_3$O$_{7-\delta}$ (YBCO). 
They observed a fourfold oscillation, in addition to 
a rather large twofold oscillation 
presumably due to the Cu-O chains specific to YBCO. 
Referring to the result based on a nonlocal GL theory,~\cite{Takanaka} 
they concluded that the observed fourfold anisotropy is 
consistent with the simple $d_{x^2-y^2}$-wave model. 
However, their interpretation strongly relies on the conjecture 
that the sign of the $\tau$-oscillation does not change 
between $H_{c1}$ and $H_{c2}$.~\cite{Takanaka2} 
In this connection, it is worth emphasizing that 
the sign of the $M_L$-oscillation {\it does} change 
between $H_{c1}$ and $H_{c2}$.~\cite{Adachi1, Kusunose2} 
Namely, the anisotropy of $M_L$ in low fields is opposite 
to what is expected from the $H_{c2}$-anisotropy. 
Thus, it is of importance to clarify the 
$\tau$-oscillation behavior in the whole field range 
from $H_{c1}$ to $H_{c2}$. 

In this work we study the 
basal-plane longitudinal magnetization $M_L$ and torque $\tau$ 
in a mixed state of layered $d$-wave superconductors 
within the quasiclassical Eilenberger formalism. 
The present paper is an extension of our previous work~\cite{Adachi1} 
on the longitudinal magnetization to the transverse component, 
and clarifies that the sign of $\tau$-oscillation does not change 
while the sign of $M_L$-oscillation changes. 
Then, it is shown that a simple interpretation 
connecting the presence or absence of sign changes 
to the anisotropy dependences of $H_{c1}$ and $H_{c2}$ is available. 
In addition, we provide the details of our analysis that were not discussed 
in our previous work. 
We use a simple $d_{x^2-y^2}$-wave model 
with the Fermi surface being isotropic within the basal plane. 
Since no detailed theoretical study on the basal-plane magnetic torque 
has been reported, 
it is meaningful to study the simple $d$-wave model. 
Of course, in an actual material, another source of complication~\cite{Adachi1} 
may be the basal-plane anisotropy of the Fermi surface. 
For YBCO, the simple $d$-wave model is appropriate for the 
qualitative study of the nonlocal effects, 
since the $d_{x^2-y^2}$-wave model 
can explain the observed configuration 
of the vortex lattice in high fields, i.e., 
a square lattice with the nearest neighbors along the nodal directions.~\cite{Brown} 

The plan of the present paper is as follows. 
In \S~2, we develop a method of solving the Eilenberger equations 
for high-$\kappa$ superconductors. 
In addition, 
we introduce the approximate solution used in our previous work,~\cite{Adachi1} 
as well as the solution given by Pesch.~\cite{Pesch} 
By comparing these three results, we show that the approximate solution 
presented in this paper gives a good result, at least with respect to 
the thermodynamics. 
This section may be skipped by those who are not so interested in the 
theoretical details. 
In \S~3, we introduce a model describing the basal-plane anisotropy 
of a layered superconductor. 
Based on the model and using the approximate solution presented in \S~2, 
we discuss the basal-plane anisotropies of 
the longitudinal magnetization and torque, 
and show that the sign of $M_L$-oscillation changes, 
while the sign of $\tau$-oscillation does not. 
At the end of \S~3, 
we give a simple interpretation on our microscopic results. 
The summary is given in in \S~4. 

\section{Quasiclassical formalism} 
We use the quasiclassical Eilenberger formalism~\cite{Eilenberger, LO} 
derived from the BCS-Gor'kov theory through an expansion 
in terms of $(p_{\rm F} \xi_0)^{-1} \ll 1$, where $p_{\rm F}$ is the Fermi momentum 
and $\xi_0=v_{\rm F}/2 \pi T_c$ with $v_{\rm F}$ and $T_c$ being the Fermi velocity 
and the transition temperature at a zero field. 
We start with the following Eilenberger equations~\cite{Eilenberger} 
for the quasiclassical Green's functions $f$, $f^{\dag}$ and $g$ 
($k_{\rm B}=\hbar=c=1$):  
\begin{eqnarray}
f( {\varepsilon_n}, \mib{p}, \mib{r})
&=&  
(2 \varepsilon_n+ {\rm i} \mib{v} \cdot \mib{\Pi} )^{-1}
\Big(
2 g( {\varepsilon_n}, \mib{p},\mib{r}) \, w_{\mib{p}} \Delta(\mib{r})
\Big), \label{Eq:Eilen1} \hspace{0.4cm} \\ 
f^{\dagger}({\varepsilon_n},\mib{p},\mib{r})
&=&
f^*({\varepsilon_n},-\mib{p},\mib{r}), \label{Eq:Eilen2} \\
g( {\varepsilon_n}, \mib{p},\mib{r})
&=&  
\sqrt{1-f( {\varepsilon_n}, \mib{p}, \mib{r})
f^{\dagger}( {\varepsilon_n}, \mib{p}, \mib{r})}. \label{Eq:Eilen3} 
\end{eqnarray}
The gap function is expressed as 
$\Delta_{\mib{p}} (\mib{r})= w_{\mib{p}} \Delta (\mib{r})$ 
where $w_{\mib{p}}$ is the pairing function with relative momentum $\mib{p}$ 
of the Cooper pair, and $\Delta(\mib{r})$ is the pair potential with 
center of mass coordinate $\mib{r}$. 
Here, $\mib{v}=v_{\rm F} \mib{{\hat p}}$ with $\mib{{\hat p}}=\mib{p}/|\mib{p}|$, 
$\mib{\Pi}=-{\rm i} \mib{\nabla}+ 2|e| \mib{A}$ with 
$\mib{A}$ being the vector potential, 
and $\varepsilon_n= 2 \pi T(n+1/2)$ 
denotes the fermionic Matsubara frequency. 
In the following, only the region $\varepsilon_n > 0$ is considered. 
We set $w_{\mib{p}}=1$ in an $s$-wave superconductor 
and $w_{\mib{p}}=\sqrt{2}(\hat{p}_x^2-\hat{p}_y^2)$ in a $d$-wave superconductor. 
The pair potential should be determined self-consistently through the 
gap equation 
\begin{equation}
  \Big( \ln (\frac{T}{T_{c}}) 
+ 2 \pi T \sum_{\varepsilon_n > 0} \varepsilon_n^{-1} \Big) \Delta
=
   {2 \pi T} \sum_{\varepsilon_n > 0}
  \langle  w_{\mib{p}}^* f \rangle, \label{Eq:Gap1} 
\end{equation}
where 
\begin{equation}
 \langle \cdots \rangle 
= \frac{\int_{\rm FS} dS (\cdots)/|\mib{v}|}
{\int_{\rm FS} dS /|\mib{v}|} 
\end{equation} 
denotes the average over the Fermi surface
with $dS$ being the area element of the Fermi surface. 

\subsection{Full solution}
In this subsection, we propose a new method of solving the Eilenberger equations 
for the vortex lattice states in extreme type-II superconductors 
with large GL parameters, $\kappa \gg 1$. 
Since this method is beyond either the approximate solution in our 
previous work~\cite{Adachi1} or Pesch's solution,~\cite{Pesch} 
the solution obtained by this method is called a ``full solution'' in this paper. 

We use the Cartesian coordinates $(x_1, x_2, x_3)$ 
constituting the right-handed system, 
and set the magnetic induction $\mib{B}$ parallel to $\mib{{\hat x}}_3 $. 
We divide the microscopic field $\mib{b}(\mib{r})$ into 
a spatially uniform part $\mib{B}$ (i.e., induction), and 
a periodic part $\delta \mib{b}(\mib{r})$ with zero mean 
within the vortex lattice cell: 
\begin{equation}
\mib{b}(\mib{r})= \mib{B}+ \delta \mib{b}({\bf r}). 
\end{equation}
Since the latter part is known to be of the order of $O(1/\kappa^2) \ll 1$ 
and negligible in high-$\kappa$ superconductors except for $H \approx H_{c1}$, 
we neglect $\delta \mib{b}$ from now on. 
To be consistent with this assumption, 
we neglect the demagnetization effect. 
The set of primitive translation vectors $\mib{m}, \mib{n}$ 
of the vortex lattice in this paper is given as 
\begin{eqnarray}
\mib{m} &=& (2 \pi /\nu) \mib{{\hat x}}_1, \\
\mib{n} &=& (2\pi \zeta/\nu) \mib{{\hat x}}_1+ \nu \mib{{\hat x}}_2, 
\end{eqnarray}
where lengths are measured in units of $r_B= (2 |e| B)^{-1/2}$ 
above and in the following. 
The real constants $\zeta$ and $\nu$ specify the form of vortex lattice. 
For a triangular vortex lattice, we set $\zeta=1/2$ and $\nu=(3 \pi^2)^{1/4}$, 
while for a square vortex lattice, we set $\zeta=1/2$ and $\nu= \sqrt{\pi}$. 
In any case, the difference in vortex configurations has a 
negligible influence on the thermodynamics 
at the level of the mean field approximation. 

We fix the gauge of the vector potential as 
\begin{equation}
  \mib{A}(\mib{r})= -Bx_2 \mib{{\hat x}}_1. 
\end{equation}  
The boundary condition for the pair potential $\Delta(\mib{r})$ is gauge dependent, 
and with the present gauge, $\Delta(\mib{r})$ 
can be expressed as the superposition of each Landau level function $\psi_N$: 
\begin{eqnarray}
  \Delta (\mib{r}) &=& \Delta_0 \sum_{N =0}^{N_{\rm max}} d_N \psi_N \label{Eq:LLEX1}, \\
  \psi_N &=& 
  \sum_{m=-\infty}^{\infty} C_m 
  \frac{ H_N(x_2+\nu m)}{\sqrt{2^N N!}} e^{-\frac{1}{2}(x_2+\nu m)^2
    -{\rm i}\nu m x_1 }, \qquad \label{Eq:LLEX2} 
\end{eqnarray}
where $\Delta_0=1.764 T_c$, 
$C_m= (\frac{\nu^2}{\pi})^{1/4} e^{- {\rm i} \pi \zeta m^2}$, 
and $H_N$ is the $N$-th Hermite polynomial. 
On the other hand, the normal component of the quasiclassical Green's function, 
$g(\varepsilon_n, \mib{p}, \mib{r})$, 
is a periodic quantity of $\mib{r}$, and it can be expanded in the lattice sums 
by Fourier transformation: 
\begin{eqnarray}
  g(\varepsilon_n, \mib{p}, \mib{r}) 
&=& \sum_{\mib{Q}} g_{\mib{Q}} (\varepsilon_n, \mib{p})
\exp ( {\rm i} \mib{Q}\cdot\mib{r} ), 
\label{Eq:FT1} \\ 
  g_{\mib{Q}}(\varepsilon_n, \mib{p}) 
&=& \Big[ g(\varepsilon_n, \mib{p},\mib{r}) 
\exp ( -{\rm i} \mib{Q}\cdot\mib{r} ) \Big]_{\rm s}, 
\label{Eq:FT2}
\end{eqnarray}
where $[\cdots]_{\rm s}$ denotes the spatial average. 
The reciprocal lattice vector $\mib{Q}$ is given by 
\begin{equation}
  \mib{Q}= m \nu \mib{{\hat x}}_1+ (-m \zeta+ n)(2 \pi / \nu) \mib{{\hat x}}_2, 
\label{Eq:Qvector}
\end{equation}
with $m$ and $n$ being integers. 
Substituting eqs.~(\ref{Eq:LLEX1}) and (\ref{Eq:FT1}) into eq.~(\ref{Eq:Eilen1}), 
we have an expression of $f$, 
\begin{eqnarray}
f &=&
2 w_{\mib{p}} \sum_{\mib{Q}, N} g_{\mib{Q}} d_N \; \beta_{\mib{Q},N},  \\
\beta_{\mib{Q}, N}
&=& \Delta_0 (2 \varepsilon_n+ {\rm i} \mib{v} \cdot \mib{\Pi} )^{-1}
(e^{{\rm i} \mib{Q}\cdot\mib{r}} \psi_N ) . 
\label{Eq:f-func}
\end{eqnarray}
The remaining task is to derive the expression of $\beta_{\mib{Q},N}$. 
Using the parameter representation 
$(2 \varepsilon_n+ {\rm i} \mib{v} \cdot \mib{\Pi} )^{-1}
= \int_0^{\infty} d \rho
e^{-(2 \varepsilon_n+ {\rm i} \mib{v} \cdot \mib{\Pi} ) \rho }$, 
we obtain 
\begin{eqnarray}
(2 \varepsilon_n+ {\rm i} \mib{v} \cdot \mib{\Pi} )^{-1}
&=&
\int_0^{\infty} d \rho \; 
e^{-2 \varepsilon_n \rho} 
e^{-{\rm i}v_1 v_2 \rho^2/r_B^2} \hspace{1cm} \nonumber \\
& & \quad \times
e^{-{\rm i}\rho v_1 \Pi_1 } e^{-{\rm i}\rho v_2 \Pi_2 } , 
\end{eqnarray}
where the operator identity $e^{A+B}= e^{-\frac{1}{2}[A,B]} e^A e^B$ was used, 
which holds if $[[A,B],A]=[[A,B],B]=0$. 
Then, by applying the identity $e^{a \partial_x} f(x)= f(x+a)$ 
for a nonsingular function $f$, 
and introducing the notation 
$\mib{s}= (\rho/r_B)(v_1 \mib{{\hat x}}_1+ v_2 \mib{{\hat x}}_2)$, 
we have 
\begin{eqnarray}
\beta_{\mib{Q}, N}
&=&
\Delta_0  \int_0^{\infty} d \rho \; e^{-2 \varepsilon_n \rho} \nonumber \\
&&\times
e^{-\frac{\rm i}{2} s_1 s_2+ {\rm i}s_1 x_2} 
\psi_N(\mib{r}- \mib{s}) e^{{\rm i}\mib{Q}(\mib{r}-\mib{s})},  
\label{Eq:beta}
\end{eqnarray}
and we can finally solve the Eilenberger equation (\ref{Eq:Eilen1}). 

To numerically obtain the self-consistent solutions for $f$, $f^\dag$, and $g$, 
we use the following procedure. 
First we give initial values for $\{ d_N \}$ and $\{ g_{\mib{Q}} \}$. 
Next we use eqs.~(\ref{Eq:f-func}) and (\ref{Eq:beta}) to obtain the new $f$, 
and eqs.~(\ref{Eq:Eilen2}), (\ref{Eq:Eilen3}) to obtain $f^\dag$ 
and $g$. 
From $g(\varepsilon_n, \mib{p},\mib{r})$, $g_{\mib{Q}}(\varepsilon_n, \mib{p})$ 
is obtained using eq.~(\ref{Eq:FT2}). 
Each $\{ d_N \}$ is determined by the following gap equation 
projected onto each Landau level: 
\begin{equation}
\bigg( \ln \Big( \frac{T}{T_c} \Big) 
+ 2 \pi T \sum_{\varepsilon_n > 0} \frac{1}{\varepsilon_n} \bigg) d_N 
=
\frac{2 \pi T}{\Delta_0} \sum_{\varepsilon_n > 0} \Big[
\psi_N^* \langle w_{\mib{p}}^* f \rangle
\Big]_{\rm s}. 
\end{equation}
Finally, we return to eq.~(\ref{Eq:Eilen1}) and repeat the same procedure until 
self-consistency is achieved. 

Once we correctly determine $f$, $f^\dag$, and $g$, 
the physical quantities can be calculated. 
In particular, 
the thermodynamic quantities can be obtained from the free energy $F/V$ 
through the thermodynamic relation. 
The expression of $\widetilde{F}/V= F/V- B^2/8 \pi$ is given as 
\begin{eqnarray}
\frac{\widetilde{F}}{V} &=& 
N(0) \Bigg[ \frac{1}{|{\cal V}|} |\Delta|^2
- 2 \pi T \sum_{\varepsilon_n > 0} \Big\langle 
\Delta^* f + \Delta f^\dag \nonumber \\
&& \quad - 
\frac{f(2 \varepsilon_n+ {\rm i}\mib{v} \cdot \mib{\Pi})f^\dag
+ f^\dag(2 \varepsilon_n+ {\rm i}\mib{v} \cdot \mib{\Pi}^*)f}
{2(1+g)} \Big\rangle
\Bigg]_{\rm s}  \label{Eq:FE1}\\
&=&
\pi T N(0) \sum_{\varepsilon_n > 0} \Bigg[ \Big\langle
\frac{(g-1) \big(f(w_{\mib{p}} \Delta)^*+f^\dag (w_{\mib{p}}\Delta)\big)}
{(g+1)}
\Big\rangle \Bigg]_{\rm s}. \nonumber \\
\label{Eq:FE2}
\end{eqnarray}
Here, $|{\cal V}|^{-1}= \ln(T/T_c)+ 2 \pi T \sum_{\varepsilon_n > 0} \varepsilon_n^{-1}$ 
is the strength of the attractive interaction. 
To proceed to the last line, 
we assumed that $f$, $f^\dag$, and $\Delta$ satisfy 
the Eilenberger equations (eqs.~(\ref{Eq:Eilen1}), (\ref{Eq:Eilen2})) 
and the self-consistent equation (eq.~(\ref{Eq:Gap1})). 

\subsection{Approximate solution}
To solve eq.~(\ref{Eq:Eilen1}) in a more numerically tractable manner, 
we introduce the approximate solution 
that was used in our previous work.~\cite{Adachi1} 
Hereafter, the solution obtained by the method presented in this subsection 
is called the ``approximate solution''. 

Let us introduce an approximation 
which is the generalization of the one 
used by Pesch,~\cite{Pesch} 
\begin{equation}
f( {\varepsilon_n}, \mib{p}, \mib{r})
\approx 
2 g( {\varepsilon_n}, \mib{p},\mib{r}) \, w_{\mib{p}} 
\Big( (2 \varepsilon_n+ {\rm i} \mib{v} \cdot \mib{\Pi} )^{-1}
\Delta(\mib{r}) \Big). 
\label{Eq:approx1}
\end{equation}
This approximation relies on the fact that the normal component of the 
quasiclassical Green's function $g$ has a weaker spatial dependence 
than the anomalous component $f$, as pointed out by 
Brandt {\it et al.}~\cite{BPT} 
We emphasize here that in contrast to Pesch's solution, 
we do not replace $g(\varepsilon_n, \mib{p}, \mib{r})$ in eq.~(\ref{Eq:approx1}) 
by its spatial average $[g(\varepsilon_n, \mib{p}, \mib{r})]_{\rm s}$, 
therefore, we can reproduce the first non-Gaussian term of the nonlocal GL 
free energy given in ref.~\citen{Adachi2}. 
As we show later, Pesch's solution can be obtained from the approximate 
solution by completely neglecting the spatial dependence of the $g$-function and 
by restricting $\Delta$ to the lowest ($N=0$) Landau level. 
Recently, Pesch's solution 
has been widely used in slightly different 
contexts.~\cite{Houghton, Dahm, Kusunose, Udagawa, Tewordt} 

To proceed with our approximation, 
it is convenient to define a new function 
$\Phi$ in a similar manner as described in ref.~\citen{Golubov} 
($\Phi$-parameterization): 
\begin{eqnarray}
  \Phi({\varepsilon_n}, \mib{p},\mib{r}) 
  &=&
  2 w_{\mib{p}} 
\Big( (2 \varepsilon_n+ {\rm i} \mib{v} \cdot \mib{\Pi} )^{-1} 
\Delta(\mib{r}) \Big). 
\label{Eq:Phi1}
\end{eqnarray}
This function offers simple representations of $f$, $f^{\dag}$, and $g$: 
\begin{eqnarray}
  f &=& g \Phi , \label{Eq:Approx1}\\
  f^{\dag} &=& g \Phi^{\dag} , \label{Eq:Approx2}\\
  g &=& \frac{1}{\sqrt{1+ \Phi \; \Phi^{\dag}}}, \label{Eq:Approx3}
\end{eqnarray}
where 
$\Phi^{\dag} ({\varepsilon_n}, \mib{p},\mib{r}) 
=
\Phi^*({\varepsilon_n}, -\mib{p},\mib{r}) $. 
Repeating essentially the same argument as described in the previous subsection, 
we obtain 
\begin{equation}
  \Phi=   2 w_{\mib{p}} 
  \sum_N \beta_{\mib{Q}=\mib{0},N} \; d_N. 
\label{Eq:Phi2}
\end{equation}
After some calculations, 
$\beta_{\mib{Q}=\mib{0}, N}$ in eq.~(\ref{Eq:Phi2}) 
can be transformed into the form used in ref.~\citen{Adachi1}, 
\begin{eqnarray}
  \beta_{\mib{Q}=\mib{0}, N}
  &=& 
  \Delta_0 \int_0^\infty d \rho e^{- 2 \varepsilon_n \rho} \alpha_N  (\mib{p}), \\
  \alpha_N (\mib{p}) &=& 
  \sum_{m= -\infty}^{\infty} C_m
  \frac{ H_N( x_2 + \nu m- {\rm Re}\, \lambda)}{\sqrt{2^N N!}} \nonumber \\
  && \times 
       {\rm e}^{-(|\lambda|^2- \lambda^2)/4}{\rm e}^{-(x_2+\nu m- \lambda)^2/2
	 -{\rm i}\nu m x_1 },
\end{eqnarray}
where $\alpha_N$ depends on $\mib{p}$ through $\lambda=(v_2+ {\rm i}v_1) \rho/r_B$. 
The numerical procedure for the above approximate method 
was explained in our previous work.~\cite{Adachi1} 

Now we show that the approximate solutions of 
eqs.~(\ref{Eq:Approx1})-(\ref{Eq:Approx3}) can reproduce the 
nonlocal GL free energy derived in ref.~\citen{Adachi2} up to the quartic term. 
We start from eqs.~(\ref{Eq:Approx1})-(\ref{Eq:Approx3}). 
Expanding $f$, $f^\dag$, and $g$ in terms of $\Delta$, we have 
\begin{eqnarray}
f &=& \Phi - \frac{1}{2} \Phi^2 \Phi^{\dag}+ \cdots, \nonumber \\
f^\dag &=& \Phi^\dag - \frac{1}{2} (\Phi^\dag)^2 \Phi+ \cdots, \\
g &=& 1- \frac{1}{2}\Phi \Phi^\dag + \cdots. \nonumber 
\end{eqnarray}
Then, we substitute these equations into eq.~(\ref{Eq:FE2}) and obtain 
\begin{eqnarray}
\frac{\widetilde{F}}{V} &=& 
N(0) \Bigg[ 
\Delta^* \bigg( \frac{1}{2|{\cal V}|} \Delta
+
\pi T \sum_{\varepsilon_n > 0} 
\Big\langle
- \Phi 
+ \frac{1}{4} \Phi^\dag (\Phi^2) 
\Big\rangle \bigg) \Bigg]_{\rm s}\nonumber \\
&&\hspace{5cm} +{\rm c.c.} 
\end{eqnarray}
up to $O(|\Delta|^4)$. 
This expression of $\widetilde{F}/V$ is equivalent to that 
derived in ref.~\citen{Adachi2} using diagrammatic perturbative calculation. 
This fact justifies the approximate solution near $H_{c2}$. 

\subsection{Pesch's solution}
It is instructive to see how the approximate solution presented above 
is connected with Pesch's solution.~\cite{Pesch} 
Pesch's solution can be derived from the approximate solution 
by imposing two additional conditions: 
(i) completely neglect the spatial dependence of the $g$-function, 
and (ii) restrict $\Delta$ to the lowest ($N=0$) Landau level. 
Below, we briefly explain how to derive Pesch's solution 
from the approximate solution 
in a two-dimensional case. 
Following Pesch (see eq.~(7) of ref.~\citen{Pesch}), 
we rewrite eq.~(\ref{Eq:Approx3}) in the form $g^2(1+\Phi \Phi^\dag)=1$.  
Keeping the above-mentioned two conditions and taking the spatial average of 
both sides of the equation, 
we obtain (cf. eq.~(27) of ref.~\citen{Dahm}) 
\begin{eqnarray}
[g(\varepsilon_n,\mib{p},\mib{r})]_{\rm s} 
&=& \Big(1+ P({\varepsilon_n, \mib{p}})\Big)^{-1/2},  \\
P({\varepsilon_n,\mib{p}})
&=& 4 |w_{\mib{p}} d_{0}|^2 
\Delta_0^2 \bigg[
\Big( \int_0^\infty d \rho_1 e^{-2 \varepsilon_n \rho_1} 
\alpha_{0}  (\mib{p}) \Big) \nonumber \\
&& \hspace{0.8cm} \times
\Big( 
\int_0^\infty d \rho_2 e^{-2 \varepsilon_n \rho_2} \alpha_{0}^* (-\mib{p}) \Big)
\bigg]_{\rm s}. 
\end{eqnarray}
After some algebra, this quantity is transformed into 
\begin{eqnarray}
P({\varepsilon_n,\mib{p}})
 &=& 4 |w_{\mib{p}} d_0|^2 \Delta_0^2 \hspace{5cm} \nonumber \\
&\times&
\int_0^\infty d \rho_1 
\int_0^\infty 
d\rho_2 e^{-2 \varepsilon_n (\rho_1+ \rho_2)
-(\rho_1+\rho_2)^2/4 \tau_B^2} \nonumber \\
&=&
4 |w_{\mib{p}} d_0|^2 (\Delta_0 \tau_B)^2  
\int_0^\infty d \rho  \rho e^{ -\rho^2/4 -2 \varepsilon_n \tau_B \rho}.   
\end{eqnarray}
Here, $\rho= (\rho_1+\rho_2)/\tau_B$ with $\tau_B= r_B/v_{\rm F}$. 
Using Dawson's integral~\cite{Dawson} 
$W({\rm i} \xi)= e^{\xi^2} {\rm erfc}(\xi)$ 
with $\xi= 2 \varepsilon_n \tau_B$, 
we finally have 
\begin{equation}
P({\varepsilon_n,\mib{p}}) 
= \Big( \frac{2 |w_{\mib{p}} \Delta_0 d_0|^2 }{\varepsilon_n^2} \Big)
\xi^2 \Big( 1- \sqrt{\pi} \xi W( {\rm i}\xi) \Big).  
\end{equation}
The above expression is equivalent to the result~\cite{Dahm} 
reformulated for a two-dimensional superconductor. 
The method of obtaining the self-consistent solution using Pesch's approximation 
is described in ref.~\citen{Dahm}.

\subsection{Validity of approximate solution} 
In this subsection, we compare 
the approximate solution in \S~2.2 with 
the full solution in \S~2.1 and Pesch's solution in \S~2.3. 
Calculations are performed in two-dimensional cases with isotropic Fermi surfaces, 
where the area element of the Fermi surface is given by 
$dS= p_{\rm F} d \varphi$ with $\varphi$ being the azimuthal angle 
of $\mib{v}$. 
First, we briefly explain our numerics. 
To perform the $\rho$-integral in the calculation of $\beta_{\mib{Q},N}$, 
we use the simple trapezoidal rule. 
In our calculation, we use the cutoff for reciprocal lattice vector 
$\mib{Q}$ in eq.~(\ref{Eq:Qvector}) as 
$-I_{Q_{\rm max}} \le m,n  \le I_{Q_{\rm max}}$; 
thus the total number of reciprocal lattice vectors equals $(2 I_{Q_{\rm max}}+1)^2$. 
For the Fermi surface average, we use Simpson's rule with an equally spaced mesh. 
For the spatial average, we use the simple trapezoidal rule, 
and we set $N_{\rm max}=12$. 

\begin{figure}[t]
\scalebox{0.5}[0.5]{\includegraphics{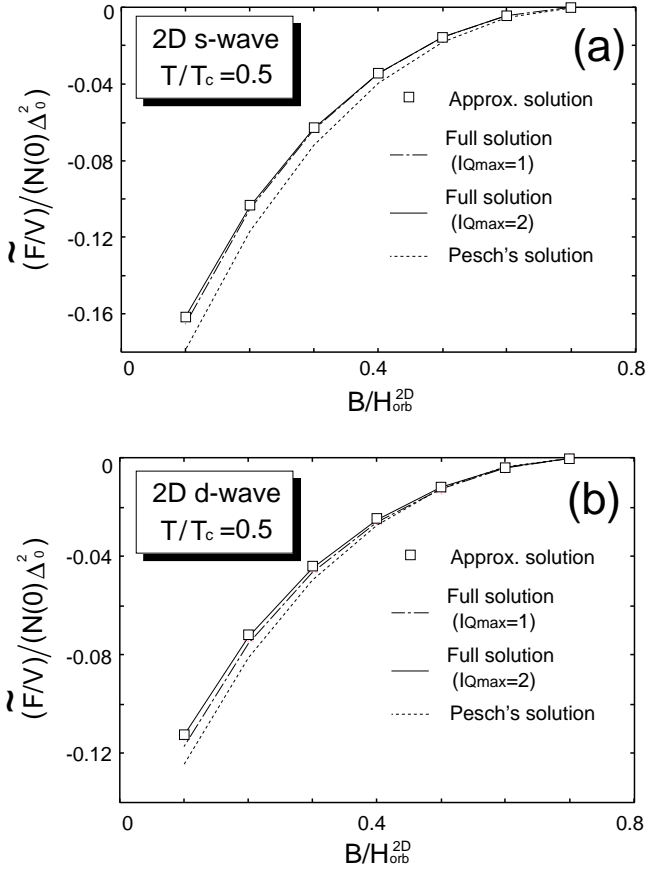}}
\caption{
Field dependences of the free energy at $T/T_c=0.5$ 
(a) in an $s$-wave superconductor, and (b) in a $d$-wave superconductor. 
The squares, dash-dotted line, solid line, and dotted line represent 
the approximate solution, 
the full solution with $I_{Q_{\rm max}}=1$, 
the full solution with $I_{Q_{\rm max}}=2$, and 
Pesch's solution, respectively. 
Here, 
$H^{\rm 2D}_{\rm orb}= 0.561 (|e|/\pi) /2 \pi \xi_0^2 \; (\xi_0= v_{\rm F}/2 \pi T_c)$ 
is the two-dimensional orbital limiting field.}
\label{Fig:FE-h_t050}
\end{figure}

Now we show the result for the free energy density 
$\widetilde{F}/V= F/V- B^2/8 \pi$. 
The magnetization can be 
derived from $\widetilde{F}/V$ through the thermodynamic relation 
$\mib{M}= - \nabla_{\mib{B}} (\widetilde{F}/V)$. 
Figure \ref{Fig:FE-h_t050} shows the field dependence of $\widetilde{F}/V$ 
for an $s$-wave superconductor and a $d$-wave superconductor. 
Here, the field $B$ is measured in units of 
the two-dimensional orbital limiting field 
$H^{\rm 2D}_{\rm orb}=
0.561 (|e|/\pi) /2 \pi \xi_0^2$. 
A triangular vortex lattice was assumed in the $s$-wave case, 
while a square vortex lattice was used in the $d$-wave case. 
We plot the results of the different methods discussed above: 
the approximate solution; 
the full solutions with $I_{Q_{\rm max}}=1$ and $2$; 
and Pesch's solution. 

As was discussed by Brandt {\it et al.},~\cite{BPT} 
the convergence by changing 
the cutoff $I_{Q_{\rm max}}$ for the reciprocal lattice vectors 
is quite excellent 
when we calculate the full solution of the Eilenberger equations. 
That is, the higher Fourier components of the $g$-function, $g_{\mib{Q} \ne \mib{0}}$, 
do not give significant contributions to the thermodynamic quantities. 
Indeed, we can see in Fig.~\ref{Fig:FE-h_t050} that 
data with $I_{Q_{\rm max}}=1$ and $2$ are almost the same. 
Moreover, data corresponding to the approximate solution and 
the full solution with $I_{Q_{\rm max}}=2$ 
cannot be distinguished on this scale. 
This means that the approximate solution provides a fairly good result, 
at least with respect to the thermodynamics. 
On the other hand, Pesch's solution seems to overestimate the 
free energy. 
This is because Pesch's solution integrates out the vortex degrees of freedom 
in advance, so that the system tends to approach the Meissner state. 

\begin{figure}[t]
 \scalebox{0.5}[0.5]{\includegraphics{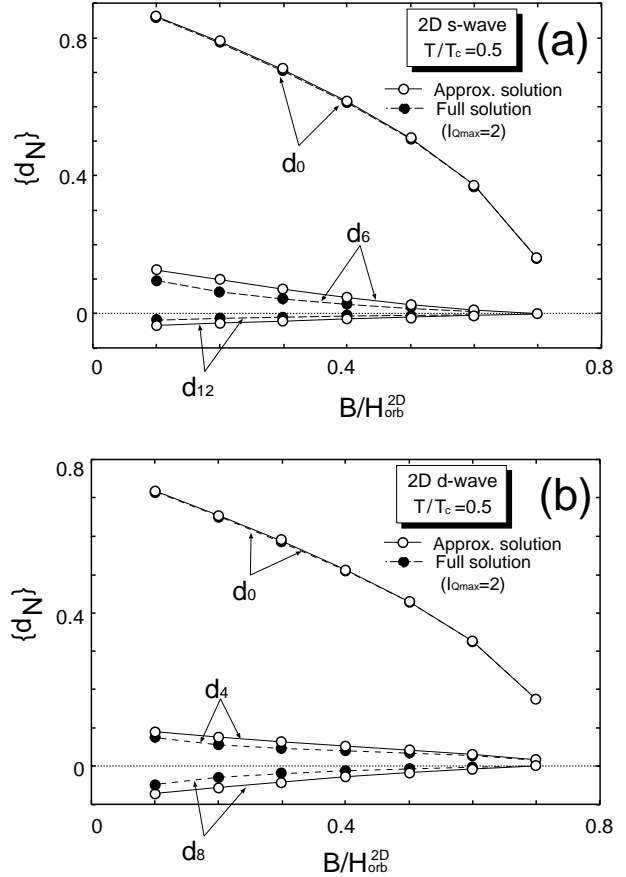}}
\caption{
Field dependence of each 
$\{ d_N \}$ at $T/T_c=0.5$ (a) in an $s$-wave superconductor, 
and (b) in a $d$-wave superconductor. 
The empty circles with the solid line represent the approximate solution,
and the filled circles with the dashed line represent the full solution 
with $I_{Q_{\rm max}}=2$. 
} 
\label{Fig:dN-h_t050}
\end{figure}

Next, we study the spatially resolved quantities. 
Figure \ref{Fig:dN-h_t050}(a) shows the field dependence of 
$\{ d_N \}$ at $T/T_c=0.5$ in an $s$-wave superconductor. 
In the figure, a slight difference appears 
between the full solution and the approximate solution. 
For the lowest Landau level component $d_0$, 
we cannot see the essential difference between 
the full solution and the approximate solution. 
On the other hand, a non-negligible difference can be seen in the field dependence of 
higher Landau level components. 
We show, in Fig. \ref{Fig:dN-h_t050}(b), the corresponding data 
in a $d$-wave superconductor. 
Since we assume a square vortex lattice in this case, 
$\{ d_N \}$ with $N=0, 4, 8, \ldots$ 
grows with decreasing field,~\cite{Mineev, Won}  
and this can be clearly seen in the figure. 
In both cases the approximate solution seems to slightly overestimate 
the amplitude of the higher Landau level components. 
Physically, the higher Landau level components describe the fine spatial 
structure of $\Delta$, and these are 
relevant to the estimation of fine structure in the local density of states, 
which is closely related 
to the differential conductance detected in the STM experiment.~\cite{Hess} 
These physics are, however, beyond the scope of this work. 
It is worth noting that the overall behavior of each $\{ d_N \}$ 
calculated by the present scheme 
shows a good correspondence to the result obtained by Watanabe 
{\it et al.}~\cite{Watanabe1} (see Fig.~2 therein), 
who solved the Eilenberger equations by a different method. 
A small difference in the $d_4$-component between our result and that in 
ref.~\citen{Watanabe1} in the $d$-wave case is considered to 
stem from the difference 
in the configuration of the vortex lattice. 
We assumed a square lattice with the nearest neighbor oriented 
along the nodal direction, 
while in ref.~\citen{Watanabe1}, 
the nearest neighbor was oriented along the antinodal direction.~\cite{Watanabe2} 

The results in this section can be summarized as follows. 
The approximate solution in \S~2.2 gives relatively good results 
as long as we consider the thermodynamics in the mixed state. 
Therefore we can use this approximation for the analysis of the basal-plane 
magnetization and torque, 
which we present in the next section. 

\section{Reversible Magnetization and Torque}

\subsection{Model for layered superconductor} 
\begin{figure}[t] 
\begin{center}
\scalebox{0.35}[0.35]{\includegraphics{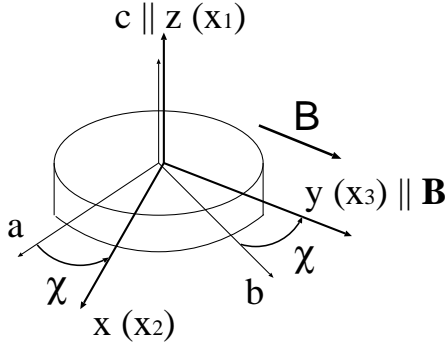}}
\end{center}
\caption{Coordinates $(x,y,z)$ and crystal axes $(a,b,c)$.
Induction $\mib{B}$ is rotated from the $b$-axis by angle $\chi$.
}
\label{Fig:setup}
\end{figure}

In this subsection, we introduce the basal-plane anisotropy 
of a layered superconductor into our treatment. 
First of all, we discuss the anisotropic model that takes into account 
a layered structure of the materials, such as cuprate superconductors. 
An example of such a procedure~\cite{Klemm, Graser} is 
the parameterization of the Fermi surface as a distorted cylinder. 
We consider a sample arrangement as shown in Fig.~\ref{Fig:setup}. 
We start with the uniaxial dispersion relation 
\begin{equation}
  \epsilon_{\mib{p}}= \frac{1}{2m}(p_x^2+ p_y^2)
  -t \cos(p_z s), 
\end{equation}
where $s$ is the interlayer spacing and $t$ is the interlayer hopping energy. 
As usual, we define the uniaxial anisotropy parameter by $\gamma=v_{\rm F}/t s$, 
where $v_{\rm F}=p_{\rm F}/m$. 
Assuming $\gamma \gg 1$ and $p_{\rm F} s \gg 1$, 
the Fermi surface is parameterized as 
\begin{equation}
  \mib{p}= 
  p_{\rm F}\mib{{\hat \rho}}  + p_z \mib{{\hat z}}, 
\end{equation}
where $\mib{{\hat \rho}}= \cos(\varphi) \mib{{\hat x}}+ \sin(\varphi)\mib{{\hat y}}$, 
$-\pi \le \varphi < \pi $ is the azimuthal angle, and $-\pi/s \le p_z < \pi/s $. 
Then, the Fermi velocity $\mib{v}= \mib{\nabla}_{\mib{p}} \epsilon_{\mib{p}}$ 
can be expressed as 
\begin{equation}
\mib{v}= v_{\rm F} \Big( \mib{{\hat \rho}}
+ \gamma^{-1} \sin(p_z s) \mib{{\hat z}} \Big), \label{Eq:Vf}
\end{equation}
and the area element $dS$ divided by $|\mib{v}|$ is given by 
$dS/|\mib{v}|= m d \varphi d p_z $. 
In this paper, we treat a simple $d_{x^2-y^2}$-wave model, and we set 
the pairing function 
\begin{equation}
w_{\mib{p}}= \sqrt{2} \cos \big(2 (\varphi+ \chi) \big), 
\end{equation}
where $\chi$ is the field angle, as shown in Fig.~\ref{Fig:setup}. 

\begin{figure}[t]
 \scalebox{1.0}[1.0]{\includegraphics{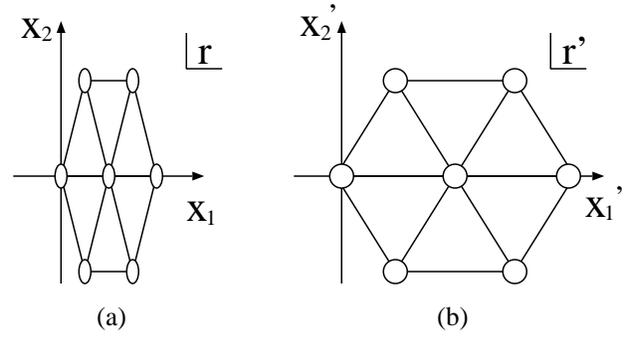}}
\caption{
Vortex lattice structure seen in (a) real space coordinates, 
and (b) distorted coordinates. 
} 
\label{Fig:coordinate}
\end{figure}

Next we define the anisotropic Landau levels 
suitable for the description of the in-plane vortex state. 
To connect the formulation in the previous section with the present 
sample arrangement (see Fig.~\ref{Fig:setup}), 
we should set $x_1=z$, $x_2=x$ and $x_3=y$. 
In the following argument, we still use the notation $(x_1,x_2,x_3)$. 
We define the raising and lowering operators ${\hat a}_\pm$ as 
\begin{equation}
{\hat a}_\pm = \frac{-1}{\sqrt{2 \eta}}\Big(\Pi_1 \pm {\rm i} \eta \Pi_2 \Big),
\end{equation}
where the real constant $\eta$ characterizes the anisotropy. 
If we choose $\eta > 1$, then the vortex state defined by the operators 
is compressed along the $x_1$-direction and stretched along the $x_2$-direction 
(see Fig.~\ref{Fig:coordinate}(a)). 
Of course, the isotropic Landau levels are defined by setting $\eta=1$. 
Now it will be convenient to introduce a new coordinate, $\mib{r'}=(x_1',x_2')$, as 
\begin{equation}
 \mib{r'}= {\hat T}\mib{r}= ({\eta}^{1/2}x_1, {\eta}^{-1/2}x_2).
\end{equation}
Then the raising-lowering operators ${\hat a}'_\pm$ and 
the resultant vortex state, expressed in the 
distorted $\mib{r'}$-frame, appear isotropic 
(see Fig.~\ref{Fig:coordinate}(b)). 
Therefore, a natural way to define the anisotropic Landau levels 
$\widetilde{\psi}(\mib{r})$ in the $\mib{r}$-frame is by 
\begin{equation}
\widetilde{\psi}_N(\mib{r}) = \psi_N({\hat T}\mib{r})
= \psi_N({\eta}^{1/2}x_1, {\eta}^{-1/2}x_2), \label{Eq:AnisoLL}
\end{equation}
where $\psi_N$ is the isotropic Landau levels defined in eq.~(\ref{Eq:LLEX2}). 
This definition is the extension of the idea used to describe 
the in-plane vortex state within the anisotropic 
lowest Landau level.~\cite{Ikeda99, Maki02, Dahm03} 
We note here that our treatment contains higher Landau level components. 
The important point in the above definition of $\widetilde{\psi}_N(\mib{r})$ 
is that if we work with the $\mib{r'}$-system in our calculation, 
the total effect of the uniaxial anisotropy can be absorbed into the 
anisotropy of the Fermi velocity represented in the distorted $\mib{r'}$-frame: 
\begin{equation}
v_1' = {\eta}^{1/2} v_1,  \qquad 
v_2' = {\eta}^{-1/2}v_2.
\end{equation}
Thus, all the formulations presented in the previous section can be valid, 
as long as we set $x_1= z$ and $x_2= x$ in the following numerical calculation. 
Although the parameter $\eta$ can be chosen arbitrarily in principle, 
the natural choice for $\eta$ is to set $\eta= \gamma$. 
This is because if we construct a linearized anisotropic GL 
equation~\cite{Gorkov} from the model eq.~(\ref{Eq:Vf}), 
then the Landau levels eq.~(\ref{Eq:AnisoLL}) with $\eta=\gamma$ 
are the solutions of it under the field parallel to the $a$- or $b$-axes. 

\subsection{Longitudinal magnetization} 
In general, magnetizations can be 
derived from $\widetilde{F}/V= F/V- B^2/8 \pi$ through the thermodynamic relation 
$\mib{M}= - \nabla_{\mib{B}} (\widetilde{F}/V)$. 
Then the longitudinal magnetization $M_L (\parallel \mib{H})$ 
is obtained by 
$M_L= - \frac{\partial}{\partial B} (\widetilde{F}/V)$, 
since the direction of the applied field $\mib{H}$ 
can be approximated by that of induction $\mib{B}$ 
in high-$\kappa$ superconductors. 
For this component, however, Klein and P\"{o}ttinger~\cite{Klein} obtained 
a more convenient formula that is the extension of the virial theorem 
derived by Doria {\it et al.}:~\cite{Doria} 
\begin{eqnarray}
 -4 \pi M_{L}
&=&
 \frac{4 \pi^2 N(0)}{B} 
T \sum_{\varepsilon_n > 0} 
\bigg[ \Big\langle 
\frac{g(f^{\dagger}(w_{\mib{p}}\Delta)+ f (w_{\mib{p}} \Delta)^*) }{1+g}
\nonumber \\
&&  \hspace{3cm} - 2 \varepsilon_n (1-g)   \Big\rangle \bigg]_{\rm s}. 
\end{eqnarray} 
In the following, we calculate the longitudinal magnetization $M_L$ 
in a layered $d$-wave superconductor under an in-plane field. 
We use the above formula for $M_L$ and assume a triangular vortex lattice 
in the $\mib{r'}$-frame. 
Since the calculation for an in-plane vortex state consumes much 
numerical resources, 
we use Simpson's rule for the spatial average as well as the Fermi surface average, 
and we set $N_{\rm max}=6$. 
\begin{figure}[t]
\scalebox{0.5}[0.5]{\includegraphics{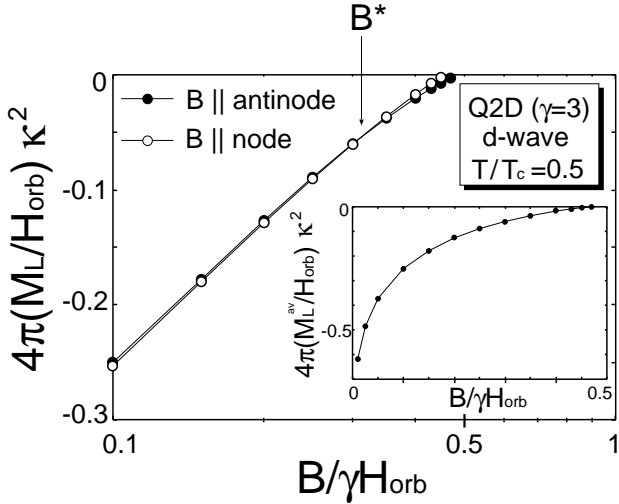}}
\caption{
Logarithmic field dependence of longitudinal magnetization $M_L$ 
in a $d$-wave superconductor. 
We set $T/T_c=0.5$ and $\gamma=3$. 
Here, $H_{\rm orb}=1.037 (|e|/\pi) /2 \pi \xi_0^2$ 
is the three-dimensional orbital limiting field in the isotropic case. 
The inset shows the linear field dependence of 
$M_L^{\rm av}=(M_L^{\mib{B} \parallel {\rm node}}- 
M_L^{\mib{B} \parallel {\rm antinode}})/2$. }
\label{Fig:magL-h_t050LOG}
\end{figure}
\begin{figure}[t]
\scalebox{0.45}[0.45]{\includegraphics{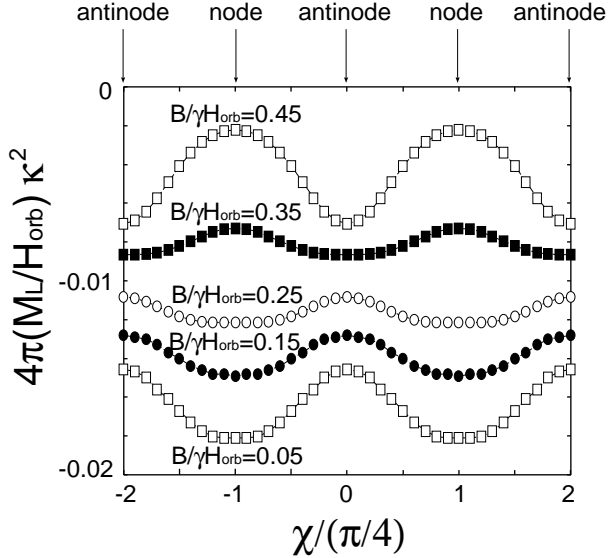}}
\caption{
Field-angle dependences of $M_L$ in a $d$-wave superconductor. 
The parameters used are the same as those in Fig.~\ref{Fig:magL-h_t050LOG}. 
The data at different $B$ are vertically shifted. }
\label{Fig:magL-chi_t050}
\end{figure}

Figure \ref{Fig:magL-h_t050LOG} shows the field dependence of $M_L$ 
in a $d$-wave superconductor. 
Here, $H_{\rm orb}=1.037 (|e|/\pi) /2 \pi \xi_0^2$ is the 
three-dimensional orbital limiting field in the isotropic case. 
We fix the temperature at $T/T_c=0.5$ and set $\gamma=3$ 
for the uniaxial anisotropy parameter. 
From the figure, we can see the London behavior $M_L \propto \ln(B_{c2}/B)$ 
in a relatively wide range of field, 
as well as the GL behavior $M_L \propto H_{c2}-B$ just below $H_{c2}$. 
This is because the GL theory is derived~\cite{Houzet} from the quasiclassical 
Eilenberger formalism through an expansion about $\Delta$, 
whereas the London theory is derived~\cite{Kogan96} 
using a phase-only (London) approximation. 
Thus we confirm that our analysis captures the essential behavior 
in the mixed state. 

Now we focus on the anisotropic properties. 
In the main panel of Fig.~\ref{Fig:magL-h_t050LOG}, 
$M_L$ for the $\mib{B} \parallel$ node (open circles) and the $\mib{B} \parallel$ 
antinode (filled circles) are shown. 
A small but clear difference in $M_L$ between the two field orientations 
is seen. 
As we examine the data more carefully, we find that the sign of the anisotropy 
changes between $H_{c1}$ and $H_{c2}$. 
Namely, $|M_L|$ for the $\mib{B} \parallel$ node is smaller near $H_{c2}$, 
while the tendency is reversed below field $B^*$. 
This is more evident in Fig.~\ref{Fig:magL-chi_t050}, 
where the field-angle dependences of $M_L$ are shown. 
Due to the inherent fourfold anisotropy of $d$-wave pairing, 
$M_L$ shows a fourfold oscillation as a function of the 
field angle $\chi$. 
Clearly, the sign of the $M_L$-oscillation changes at field 
$B^*/\gamma H_{\rm orb} \approx 0.3$ between $H_{c1}$ and $H_{c2}$. 
Thus, the sign change of $M_L$-oscillation can occur even in a simple 
$d$-wave superconductor.

\subsection{Torque}
Next we discuss the transverse component of the magnetization, or the torque. 
The transverse magnetization $M_T$ 
is given by $M_T= -B^{-1} \frac{d}{d \chi}(\widetilde{F}/V)$, where $\chi$ 
is the field angle shown in Fig.~\ref{Fig:setup}. 
In experiments, this quantity is mostly measured as a torque density, 
$\mib{\tau}/V= \mib{M} \times \mib{B}$ or $\tau/V=- \frac{d}{d \chi}(\widetilde{F}/V)$. 
To obtain our numerical data of $\tau$, we first calculate the $\chi$-dependence 
of $\widetilde{F}/V$, then perform a polynomial interpolation, 
and calculate $\tau$ using the above relation. 

\begin{figure}[t]
\scalebox{0.48}[0.48]{\includegraphics{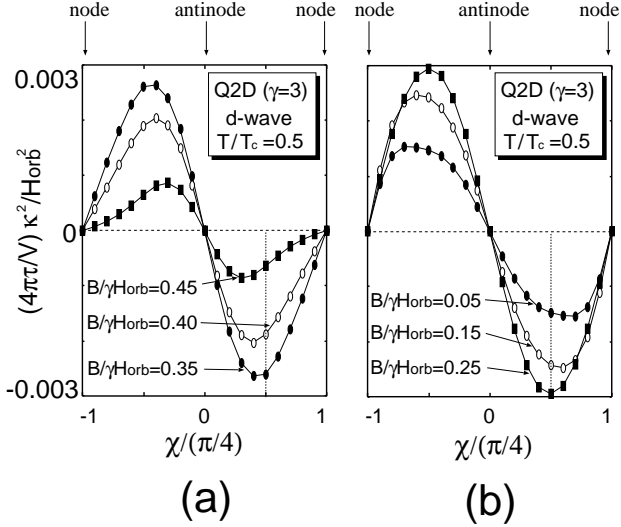}}
\caption{
Field-angle dependences of $\tau$ in $d$-wave superconductor. 
The parameters used are the same as those in Fig.~\ref{Fig:magL-h_t050LOG}. 
Data in (a) high fields and (b) low fields are shown. 
}
\label{Fig:tau-chi_d1t050}
\end{figure}

Figure \ref{Fig:tau-chi_d1t050} shows the field-angle dependences of torque 
$\tau$ in a $d$-wave superconductor at $T/T_c=0.5$. 
As in the case of longitudinal magnetization, $\tau$ shows 
a fourfold oscillation as a function of $\chi$. 
When the field is lowered from $H_{c2}$ 
($\approx 0.475 \gamma H_{\rm orb}$ in this case), 
the oscillation amplitude first increases up to $B/\gamma H_{\rm orb} \approx 0.3$, 
as seen in Fig.~\ref{Fig:tau-chi_d1t050}(a), 
whereas it starts to decrease at lower fields, 
as seen in Fig.~\ref{Fig:tau-chi_d1t050}(b). 

\begin{figure}[t]
\scalebox{0.45}[0.45]{\includegraphics{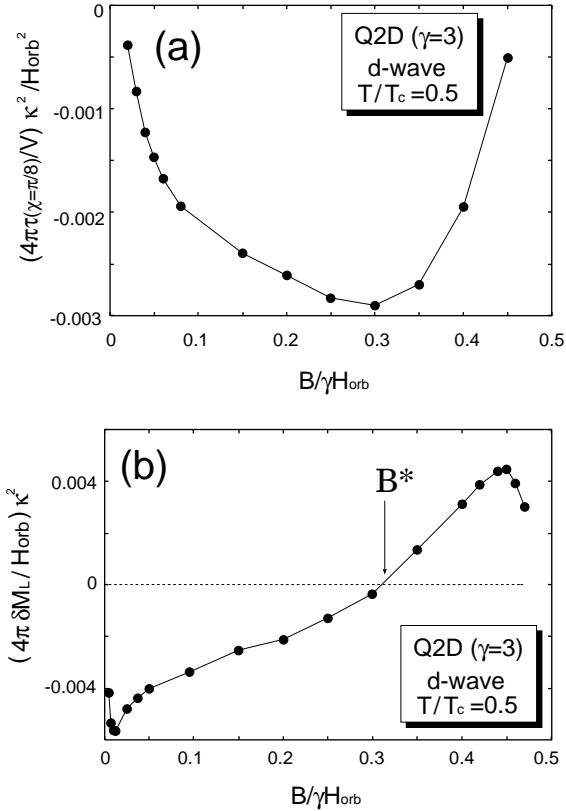}}
\caption{
Field dependence of (a) torque and (b) $\delta M_L= M_L(\chi=\pi/4)-M_L(\chi=0)$ 
in $d$-wave superconductor. 
The parameters used are the same as those in Fig.~\ref{Fig:magL-h_t050LOG}. 
}
\label{Fig:tau-Dmag_d1t050}
\end{figure}

To see the field dependence of the oscillation amplitude, 
we plot, in Fig.~\ref{Fig:tau-Dmag_d1t050}(a), 
the field dependence of $\tau$ at a fixed angle $\chi=\pi/8$. 
In the figure, we can see that the sign of the oscillation does not change 
between $H_{c1}$ and $H_{c2}$. 
Although the accuracy of our numerical solution breaks down near $B= 0$ 
or $H=H_{c1}$, 
we believe that this behavior of $\tau$ is not changed qualitatively 
by including the screening effect. 
Later, we will give a theoretical interpretation of the phenomena. 
It is worth noting that the behavior is in contrast to that in the case of 
longitudinal magnetization $M_L$, 
where the sign of the oscillation amplitude is reversed at field $B^*$ 
between $H_{c1}$ and $H_{c2}$. 
For comparison, we show, in Fig.~\ref{Fig:tau-Dmag_d1t050}(b), 
the corresponding field dependence of $\delta M_L \equiv M_L(\chi=\pi/4)-M_L(\chi=0)$. 
It is interesting that 
near field $B^*$ where the sign of $\delta M_L$ is changed, 
the field dependence of the torque is also changed from 
a decreasing one to an increasing one. 
This coincidence is explained in the next subsection. 

\subsection{Interpretation of results}
\begin{figure}[t]
\scalebox{0.45}[0.45]{\includegraphics{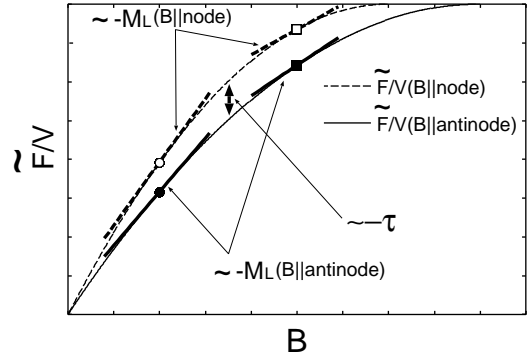}}
\caption{
Schematic field dependences of $\widetilde{F}/V$ 
for $\mib{B} \parallel$ antinode (solid line) and $\mib{B} \parallel$ 
node (dashed line). 
}
\label{Fig:FE-h_schem}
\end{figure}

We discuss the interpretation of the results that 
the sign of $M_L$-oscillation changes, while the sign of $\tau$-oscillation 
does not. 
For this purpose, it is useful to consider the anisotropy of the free energy 
$\widetilde{F}(\chi)$. 
On the basis of our numerical results, 
we schematically plot, in Fig.~\ref{Fig:FE-h_schem}, 
the field dependences of $\widetilde{F}(\chi)$ 
in a $d$-wave superconductor, corresponding to the two different field orientations 
($\mib{B} \parallel$ node and antinode). 
Here, the magnitude of anisotropy is enlarged for clarity. 
It is important to note that 
the two curves do not cross, and 
$\widetilde{F}/V$ is a convex function of $B$. 
This can be understood as follows. 

Firstly, it is important to note the relationship 
between the anisotropies of coherence length and 
penetration depth.~\cite{Gorkov, Kogan02} 
Extending this knowledge, 
we obtain the anisotropy relation between 
$H_{c1}$ and $H_{c2}$:~\cite{Adachi1} 
\begin{equation}
\frac{H_{c1}(\chi=0)}{H_{c1}(\chi=\pi/4)} \approx 
\frac{H_{c2}(\chi=\pi/4)}{H_{c2}(\chi=0)}. 
\label{Eq:Hc2Hc1}
\end{equation}
This implies that if $H_{c2}$ in a certain direction is larger than in 
other directions, 
then $H_{c1}$ in that direction is smaller. 
This can be explained in an alternative way. 
Recall that $H_{c1} = (\ln \kappa/\sqrt{2}\kappa) H_c $ 
and $H_{c2} = (\sqrt{2} \kappa) H_c $,~\cite{Fetter} where 
$H_c$ is the thermodynamic critical field. 
By definition, $H_c$ does not depend on the field orientation, 
and we can ascribe the anisotropy to $\kappa$. 
From these expressions, it is obvious that $H_{c1}$ and $H_{c2}$ 
have opposite tendencies of anisotropies in the high-$\kappa$ case. 

Next we consider the anisotropy of $\widetilde{F}/V$ 
using the analytical results.~\cite{deGennes} 
\begin{equation}
\widetilde{F}/V \approx 
\left\{ 
\begin{array}{lr}
  -(H_{c2}-B)^2/16\pi \kappa^2 & (H \approx H_{c2}) \\
 F(B=0)/V + B H_{c1}/4 \pi & (H \approx H_{c1}) \\
\end{array} \right. 
\label{Eq:GL-London}
\end{equation}
Near $H_{c1}$, the anisotropy of $\widetilde{F}/V$ 
is determined by that of $H_{c1}$, 
while near $H_{c2}$, the anisotropy of $\widetilde{F}/V$
is governed by that of $H_{c2}$. 
Using eqs.~(\ref{Eq:Hc2Hc1}) and (\ref{Eq:GL-London}), 
the free energy anisotropy shown in Fig.~\ref{Fig:FE-h_schem} 
can be confirmed analytically near $H_{c1}$ and $H_{c2}$. 
In the intermediate field region, 
it is very difficult to imagine that 
the two curves $\widetilde{F}(\chi=0)$ and $\widetilde{F}(\chi=\pi/4)$ cross 
at some field, i.e., the free energy becomes $\chi$-independent. 
Otherwise, the $d$-wave superconductor appears effectively isotropic 
at this particular field. 
In this way, the free energy anisotropy shown in Fig.~\ref{Fig:FE-h_schem} 
can be understood. 
Moreover, we can show that $\widetilde{F}/V$ is a convex function of $B$ 
if we correctly include the vortex interaction energy.~\cite{deGennes}  

Now let us first check the sign reversal of $M_L$-oscillation 
using Fig.~\ref{Fig:FE-h_schem}. 
$M_L$ is given by $M_L = -\frac{\partial}{\partial B} (\widetilde{F}/V) $, 
i.e., the negative slope of 
the field dependence of $\widetilde{F}/V$. 
Then it is obvious from Fig.~\ref{Fig:FE-h_schem} that 
$M_L$ near $H_{c2}$ is larger for the $\mib{B} \parallel$ node, 
while near $H_{c1}$, this tendency is reversed. 
This explains the sign change of $M_L$-oscillation. 
On the other hand, 
the in-plane torque $\tau= - \frac{d}{d \chi} \widetilde{F}$ 
in the simple $d$-wave case is roughly estimated by 
comparing $\widetilde{F}(\chi)$ in the two characteristic directions, 
$\chi=0$ ($\mib{B} \parallel $ antinode)
and $\chi=\pi/4$ ($\mib{B} \parallel$ node). 
Namely, we have the simple relation 
$\tau \sim -\delta \widetilde{F}= \widetilde{F}(\mib{B} \parallel {\rm node})
- \widetilde{F}(\mib{B} \parallel {\rm antinode})$. 
As is seen in Fig.~\ref{Fig:FE-h_schem}, 
the sign of this quantity is not changed between $H_{c1}$ and $H_{c2}$. 
Thus we immediately conclude that 
the sign of $\tau$-oscillation is not changed. 
In the same way, we can see that 
$-\tau \sim \delta \tilde{F}$ is maximum near field $B^*$ where 
the difference in the slope of the two curves in Fig.~\ref{Fig:FE-h_schem}, 
i.e.,  
$-\delta M_L = 
\frac{\partial \widetilde{F}/V}{\partial B} (\mib{B} \parallel {\rm node}) 
-\frac{\partial \widetilde{F}/V}{\partial B} (\mib{B} \parallel {\rm antinode})$, 
changes its sign. 
In this way, 
we can understand why near $B^*$, where 
the sign of $\delta M_L$ is changed, 
the field dependence of the torque is also changed from 
a decreasing one to an increasing one. 

We end this section with some comments on the effect of Fermi surface anisotropies. 
This effect can be studied by introducing parameter $\beta$, that describes the 
in-plane anisotropy of the Fermi surfaces, 
in a similar manner as in ref.~\citen{Adachi1}. 
We performed a calculation for when this Fermi surface anisotropy essentially 
competes with the gap anisotropy, 
and confirmed that the observed tendency of anisotropy is reversed at around 
$|\beta|=1.0$ (similar to the difference in Figs. 3(a) and 3(b) 
of ref.~\citen{Adachi1}). 
As the reader may note, however, the above argument 
for the sign change of $M_L$-oscillation 
and the sign preservation of $\tau$-oscillation 
is independent of the source of anisotropy. 
The key is the use of eq.~(\ref{Eq:Hc2Hc1}), i.e., 
the fact that $H_{c1}$ and $H_{c2}$ have the opposite tendencies of anisotropy, 
and this relation holds irrespective of the source of anisotropy. 
Thus, even if the system is strongly affected by 
the Fermi surface anisotropy, the statement 
that the direction with maximal $H_{c2}$ is always 
energetically stable in the whole field range from $H_{c1}$ to $H_{c2}$ 
is valid. 
The importance of this fact is discussed in the next section. 
Repeating the same argument as in the previous paragraph, 
we can say that the sign change of $M_L$-oscillation 
and the sign preservation of $\tau$-oscillation 
are universal. 

\section{Conclusion}
In this paper, we proposed a new method of 
obtaining the ``full solution'' 
of the quasiclassical Eilenberger equations 
for high-$\kappa$ superconductors, in addition to 
introducing a reasonable ``approximate solution'' to 
examine thermodynamic quantities. 
The approximate solution discussed here is a natural extension of the one 
developed by Pesch~\cite{Pesch} and which has been used recently 
by several authors.~\cite{Houghton, Dahm, Kusunose, Udagawa, Tewordt} 
We compared the approximate solution with the full solution or with Pesch's solution, 
and showed that our approximate method gives an excellent result, 
at least with respect to the thermodynamic quantities. 
Recently, we have extended the approximate solution to the 
case including Pauli paramagnetism.~\cite{Adachi3} 

We then applied our approach to the description of the in-plane magnetic 
anisotropies in a layered $d$-wave superconductor such as high-$T_c$ cuprate. 
We clarified that the sign of the longitudinal magnetization 
oscillation changes between $H_{c1}$ and $H_{c2}$, 
while the sign of the torque oscillation does not. 
Based on the analytical expressions of the free energy and the 
anisotropy relation between $H_{c1}$ and $H_{c2}$, 
we showed that there is a close relationship between 
$M_L$-oscillation and $\tau$-oscillation. 

Furthermore, we revealed that the basal-plane torque data always indicate 
the maximal $H_{c2}$-direction. 
The implication of the present result is that 
the basal-plane $H_{c2}$ anisotropies can be studied through torque measurement 
at low fields, 
even in materials with strong fluctuations, such as high-$T_c$ cuprates and 
organic superconductors, 
where the true $H_{c2}$ cannot be determined experimentally. 
Moreover, we can detect the node position 
if we limit ourselves to a $d$-wave superconductor possessing a 
less anisotropic Fermi surface. 
For example, in light of our results, 
the torque data for YBCO~\cite{Ishida} are consistent with $d_{x^2-y^2}$-pairing. 
In this respect, it will be interesting to measure the basal-plane torque 
in the $d$-wave superconductor $\kappa$-(ET)$_2$Cu(NCS)$_2$, 
since the node position concluded from the earlier 
theoretical analysis~\cite{Schmalian, Kino, Moriya} 
and the recent experiments~\cite{Arai, Izawa01} are controversial.~\cite{Kuroki} 
Also of interest is the measurement of the basal-plane magnetic anisotropies 
in the $d$-wave superconductor CeCoIn$_5$, 
since there again is a controversy concerning the node position.~\cite{Aoki, Izawa2}

\acknowledgements
We are grateful to T. Sakakibara, R. Ikeda, Y. Matsuda, and K. Yamada for 
helpful discussions. 
One of the authors (H. A.) would like to thank K. Watanabe for useful comments, 
and E. Ohmichi, W. Pogosov, and T. K. Ghosh for discussions.


\end{document}